\documentclass[5p,times,twocolumn]{elsarticle}
\biboptions{sort&compress}
\usepackage[T1]{fontenc}
\usepackage[utf8]{inputenc}
\usepackage{microtype}
\usepackage{hyperref}
\usepackage{graphicx}
\usepackage{amsmath,amssymb,amsfonts,mathtools}
\usepackage{bm}
\usepackage{array}
\usepackage{booktabs}
\usepackage{multirow}
\usepackage{xcolor}
\usepackage{listings}
\usepackage{algorithm}
\usepackage{algpseudocode}
\usepackage{enumitem}
\usepackage{url}
\usepackage{hyperref}
\usepackage{caption}
\usepackage{subcaption}
\usepackage{tikz}
\usepackage{float}

\usetikzlibrary{arrows.meta,positioning,fit,shapes.geometric,calc}

\hypersetup{
  colorlinks=true,
  linkcolor=blue!55!black,
  citecolor=blue!55!black,
  urlcolor=blue!55!black,
  pdfauthor={Rishabh Gupta, Kangkan Goswami, Suraj Prasad, Raghunath Sahoo},
  pdftitle={MAGE-HEP: Monte Carlo Analysis and Graphical Environment for High-Energy Physics}
}

\newcommand{\pT}{p_{\mathrm{T}}}
\newcommand{\Nch}{N_{\mathrm{ch}}}

\newcommand{\mgp}{\texttt{.mgp}}

\newcommand{\code}[1]{\texttt{#1}}

\lstdefinestyle{papercode}{
  basicstyle=\ttfamily\fontsize{7.8pt}{8.4pt}\selectfont,
  columns=fullflexible,
  keepspaces=true,
  breaklines=true,
  showstringspaces=false,
  frame=none,
  backgroundcolor=\color{white},
  xleftmargin=0pt,
  xrightmargin=0pt,
  aboveskip=0.25em,
  belowskip=0.15em,
  numbers=none
}

\tikzset{
  magebox/.style={
    draw=black!45,
    rounded corners=2pt,
    align=center,
    minimum height=7mm,
    inner sep=4pt,
    fill=white
  },
  mageblue/.style={magebox, fill=blue!7, draw=blue!45!black},
  magegreen/.style={magebox, fill=green!8, draw=green!35!black},
  mageorange/.style={magebox, fill=orange!10, draw=orange!60!black},
  magegray/.style={magebox, fill=black!4, draw=black!45},
  magearrow/.style={-Latex, line width=0.45pt, draw=black!70}
}

\begin{document}

\begin{frontmatter}

\title{\textsc{MAGE-HEP}: Monte Carlo Analysis and Graphical Environment for High-Energy Physics}

\author[inst1]{Rishabh Gupta}
\author[inst1]{Kangkan Goswami}
\author[inst1,inst2]{Suraj Prasad}
\author[inst1]{Raghunath Sahoo\corref{cor1}}

\ead{Raghunath.Sahoo@cern.ch}
\cortext[cor1]{Corresponding author}

\address[inst1]{Department of Physics, Indian Institute of Technology Indore,
Simrol, Indore 453552, India}

\address[inst2]{HUN-REN Wigner Research Centre for Physics,
29--33 Konkoly-Thege Mikl\'os Str., H-1121 Budapest, Hungary}

\begin{abstract}
Monte Carlo event generators are central to high-energy physics analysis. However, workflows based on handwritten scripts can be difficult to reuse, modify, and reproduce when multiple Monte Carlo models, tune variations, run variations, and output formats are involved. We present MAGE-HEP, short for Monte Carlo Analysis and Graphical Environment for High-Energy Physics, a Graphical User Interface (GUI) driven workflow environment for reproducible Monte Carlo-based analyses in high-energy physics. MAGE-HEP organizes analysis workflows through a project-study-run hierarchy. The project stores the workspace, the study stores the reusable analysis context, and each run represents a controlled execution of that context.

The MAGE-HEP Node API provides the analysis-building layer for defining generator configurations, observables, selections, output rules, and generated C++/ROOT analysis code. A study context can be inspected, reused, or exported as a \texttt{.mcx} context bundle, while the project state can be exported as a portable \texttt{.mgp} bundle. The current beta implementation validates the core idea using a PYTHIA8 and ROOT workflow. It includes background execution, manifest-based run tracking, live ROOT inspection, and particle-table summaries for supported output layouts. This paper describes the architecture, workflow, and current beta implementation of MAGE-HEP.
\end{abstract}


\end{frontmatter}
\section{Introduction}

Monte Carlo (MC) event generators are essential tools in High Energy Physics, used to simulate particle collisions and the subsequent evolution of final-state particles in collider experiments. They provide a theoretical-to-experimental bridge by modeling the full collision process. Generators such as PYTHIA, AMPT, HERWIG, EPOS, etc., are extensively used to interpret collider data and test predictions of Quantum Chromodynamics (QCD). In experimental analyses, MC generators serve several key purposes. They are used to estimate detector acceptance and reconstruction efficiencies, optimize event selection strategies, evaluate backgrounds, and study systematic uncertainties. This is primarily done when simulated events are passed through detector simulation frameworks to mimic the detector response, allowing direct comparison with measured data~\cite{pythia8,Sjostrand:2014zea,Bierlich:2022pfr,Campbell:2022qmc,HSFPhysicsEventGeneratorWG:2020gxw,Boyle:2022cvo,Prasad:2025yfj}. Recent community reports and generator-focused studies emphasize that event generators are no longer isolated programs. They are part of a larger computational ecosystem involving validation, tuning, standard interfaces, software sustainability, training, preservation, and reproducible use in experimental and phenomenological workflows~\cite{pythiafacility,Campbell:2022qmc,HSFPhysicsEventGeneratorWG:2020gxw,HEPSoftwareFoundation:2017ggl,futurehepsoftware,HSF:2025software,Prasad:2025yfj}.

In practice, the first analysis workflow for a new user often seems much less structured. A simple generator-level study usually requires the user to configure the generator, write the event loop, write the particle loop, apply event-level and particle-level selections, define histograms, choose output names, manage random seeds, split large runs into smaller jobs, merge outputs, and record which settings produced which result. These operations are standard for experienced users, but they can make the first Monte Carlo analysis difficult to organize and reproduce. The problem becomes more apparent when the same analysis must be repeated with different parameters across several MC models.

The high-energy physics software ecosystem continues to evolve with new analysis needs, computational challenges, and reproducibility requirements, and several mature tools already address important parts of this landscape. ROOT is widely used for storing and analyzing high-energy physics outputs~\cite{root}. HepMC and HepMC3 provide standard event-record formats for generator-level information exchange~\cite{Dobbs:2001ck,Buckley:2019xhk}. Rivet supports particle-level analysis preservation and generator validation~\cite{rivet,Bierlich:2019rhm}, while Professor is used for systematic event-generator tuning~\cite{Buckley:2009bj}. MadAnalysis~5 provides a user-friendly framework for collider phenomenology and recasting studies~\cite{Conte:2012fm}. CutLang and the Analysis Description Language provide human-readable ways to describe analysis logic~\cite{cutlang,adl,Prosper:2021umk}. REANA and CERN-style preservation systems support reusable computational workflows and analysis preservation~\cite{reana,cap,Maguire:2017ypu}. These frameworks constitute important components of the HEP software ecosystem.

MAGE-HEP, the Monte Carlo Analysis and Graphical Environment for High-Energy Physics, is developed as a context-driven workflow layer on top of existing Monte Carlo and analysis tools. It is designed to help early-stage users build reproducible generator-level workflows, while also allowing experienced users to manage repeated analyses, parameter variations, logs, software versions, generated code, and output metadata without manually maintaining separate scripts. The framework targets the stage where a physics idea must be translated into generator configurations, event loops, particle selections, ROOT objects, and output files. Typical examples include particle spectra, multiplicity-dependent observables, and acceptance-dependent distributions. Existing workflow-preservation tools can preserve and rerun structured workflows, but they do not usually define the generator-level analysis logic itself as a reusable object. MAGE-HEP addresses this by introducing a reusable study context and organizing each analysis in a project-study-run hierarchy, where the project stores the workspace, the study stores the reusable analysis definition, and each run represents one controlled execution with its own seed, parameters, logs, status files, output paths, and generated files.

The current beta version validates this idea with a PYTHIA8 and ROOT workflow, which serves as the starting implementation. It provides a Qt/C++ workspace, manifest-based project organization, background execution, generated C++/ROOT analysis code, ROOT output production, and selected visual validation tools. The MAGE-HEP Node API forms the analysis-building layer. It allows reusable logic tools to expose user-defined editable parameters and output sockets, describing the main stages of the analysis from event generation to ROOT output. The selected analysis context can then be exported as a MAGE-HEP context file and translated into a C++ analysis program. This design keeps the generated code visible to the user, avoiding treating the framework as a black box.

A key part of the MAGE-HEP design is that quantities produced at different stages of the analysis can be explicitly connected through data sockets and written to ROOT according to stored output rules. This preserves the relation between the physics quantity, the analysis stage, the output object, and the ROOT layout as part of the study context. MAGE-HEP also records run manifests, status files, logs, generated source files, ROOT layout descriptions, parameter values, and output paths, so that the ROOT file is not treated as the only reproducibility record. This aligns with the broader direction in HEP software, where reproducibility, metadata, provenance, and long-term usability are treated as part of the analysis workflow~\cite{HEPSoftwareFoundation:2017ggl,Davies:2022bfz, fair,Feickert:2023ajm}. The present implementation validates the proposed workflow model, code-generation strategy, and execution framework. Future versions can extend the same structure toward additional generator backends, HepMC-based event records, broader ROOT and YODA output layouts, real-data import, MC-to-data comparison, tuning workflows, and ML-assisted analysis tools~\cite{Dobbs:2001ck,Buckley:2019xhk,yoda,Maguire:2017ypu,Buckley:2014ana,delphes,geant4}.

The paper is organized as follows. Section~\ref{sec:design-goals} presents the main design goals of MAGE-HEP. Section~\ref{sec:workflow} describes the project-study-run workflow and the organization of projects, studies, runs, and execution metadata. Section~\ref{sec:NodeAPI} introduces the MAGE-HEP Node API and explains how reusable study contexts are defined, generated, exported, and reused. Section~\ref{sec:root-rules} discusses the ROOT builder and the current \textit{save-rule} structure used to map internal stores to ROOT objects. Section~\ref{sec:gui} describes the graphical interface and the background execution layer used for project creation, run launching, log inspection, and output viewing. Section~\ref{sec:output} presents a representative PYTHIA8/ROOT output generated with the current beta implementation. Finally, Section~\ref{sec:conclusion} summarizes the present status of MAGE-HEP and outlines the planned extensions.

\section{Design Goals}
\label{sec:design-goals}

MAGE-HEP is structured around five design goals, each motivated by a practical need to develop a more coherent and reproducible Monte Carlo analysis workflow. The current beta implementation does not claim to fully realize all of these goals. Instead, it introduces the conceptual and technical foundations required to address them systematically. These goals define the main directions in which MAGE-HEP is being developed, while the present implementation provides an extensible foundation for future improvements.

\subsection{User Interface}

The primary goal of the MAGE-HEP user interface is to provide a user-friendly environment where users can create a project, define or import a study, launch one or more runs, monitor each run's progress, inspect logs, and view supported ROOT plots without manually inspecting individual output files. This interface can serve as a visual aid for students and new users entering Monte Carlo analysis workflows, while also helping experienced users manage multiple analysis runs without manually organizing each workflow directory.

In the current beta implementation, live ROOT plots and particle tables are included as diagnostic visual tools for selected supported analysis formats. They are not intended to define the full output model of MAGE-HEP, but are used to test whether the internal study context is correctly translated during code generation and whether the generated analysis can parse the expected ROOT variables. This restriction keeps the current version reliable and testable while the core project-study-run workflow is being validated.

\subsection{Reproducibility}

Maintaining a consistent, reproducible analysis workflow is inherently challenging when analysis scripts, generator settings, plots, and output files are stored separately. MAGE-HEP treats each project as a reproducible package. A project created in MAGE-HEP can be serialized to a portable \texttt{.mgp} bundle. This bundle captures the generator configurations, studies, different run states for each study, their metadata, selected parameters, generated source files, output layout descriptions, parallel execution settings, and the tool versions used.

In the current implementation, the \texttt{.mgp} bundle stores the lightweight files required to reconstruct the workflow. Large data files, such as ROOT output files, are excluded from the current storage policy and stored externally to avoid large project bundles while preserving the relationship between the study context, run parameters, and produced outputs. In future versions, the MAGE-HEP project bundle may also be integrated with external reproducibility and workflow execution platforms, such as REANA~\cite{reana}. This would allow MAGE-HEP projects to be archived or executed within larger analysis preservation environments, while retaining the project-study-run structure used by the MAGE-HEP interface.

\subsection{Parallel Processing}

Analyses using a specific Monte Carlo can be repeated multiple times with different seeds, event counts, acceptance criteria, binning, parameter changes, or multiple parallel processes to save time. MAGE-HEP therefore separates each analysis into a `run' under the same `study,' separating the study-level definition from the run-level execution. Multiple runs can be created within the same study and executed in parallel across available CPU cores. The GUI does not execute long jobs directly. Instead, the execution job in MAGE-HEP is delegated to a backend module called \texttt{mage-daemon}. Each run is executed within a separate \texttt{tmux} session, enabling independent logging, multi-threading, and background execution~\cite{tmux}.

\subsection{Canonical Output and Documentation}

The output of a Monte Carlo analysis should not be limited to a ROOT file alone. A reproducible workflow also requires metadata describing how the output was produced. For this reason, one of the design goals of MAGE-HEP is to record run manifests, status files, logs, ROOT layout descriptions, generated source files, and parameter values together with the analysis output. Here, a run manifest is a machine-readable JSON file that stores the run configuration, seed information, execution mode, output paths, generated files, timestamps, and run status. In the current version, the analysis output is restricted to ROOT-based objects, and the ROOT layout is defined by \textit{save-rule}s stored in the internal study context. A save-rule is a set of instructions that specifies which quantity is written, which variables are used as axes, and how the object is organized inside the ROOT file. In future versions, this output layer can be extended to support additional standard event-record formats, such as HepMC~\cite{Dobbs:2001ck}, allowing MAGE-HEP to store not only derived histograms but also generator-level event information in a more interoperable form.

MAGE-HEP is not intended to be limited to generating a separate standard C++ file for each analysis. Instead, it defines a `study context' in which the user can describe the logic for the observables and other parameters required in an analysis. From this description, MAGE-HEP derives an internal context graph that allows the observables to be connected through reusable nodes, as described in Sec.~\ref{sec:NodeAPI}. The generated analysis then follows a canonical output template, making the workflow easier to inspect, reproduce, and extend.

\subsection{Cross-Generator Scripting}

The current implementation of MAGE-HEP focuses on PYTHIA8 and ROOT-based outputs. However, the workflow is designed so that the event source, analysis context, output rules, and run execution are represented separately. This separation is intentional, as it provides a path to support additional generators and output formats in later versions.

In future versions, MAGE-HEP can be extended to interoperate with existing HEP tools, such as Rivet toolkit~\cite{rivet}, for analysis preservation and validation, and with HepMC-based event records for standard generator-level data exchange. Such extensions would allow MAGE-HEP to connect its project-study-run structure to more widely used HEP analysis and event record formats. Cross-generator and external-tool support is treated as a design direction rather than a completed feature in the current beta version. The present implementation focuses on validating the PYTHIA8 and ROOT workflow.

\section{Overview of MAGE-HEP Workflow}
\label{sec:workflow}

This section presents the overall workflow of MAGE-HEP. In the current version, MAGE-HEP follows a project-study-run hierarchy, which separates the analysis workspace, reusable study context, and individual run execution. The project level organizes the complete analysis workspace, the study level stores the reusable analysis definition, and the run level executes controlled variations of that context. The processing of each layer, from project creation to execution and output inspection, is described in the following discussion.
\subsection{Project, Study, and Run}

The top-level object is the project. A project represents a complete analysis store and may contain one or more studies. A study stores a reusable analysis context. This context contains the generator configuration, selected analysis recipe, observable definitions, output rules, mutable parameters, generated analysis source files, and the versions of the tools used.

A run is a concrete execution of a study. It does not redefine the full analysis. Instead, it inherits the study-level context and applies a set of restricted run-specific changes. These changes may include the number of events, the random seed, binning, particle species, or other parameters used in the corresponding study. The workflow is summarized in Fig.~\ref{fig:project-study-run}.

\begin{center}
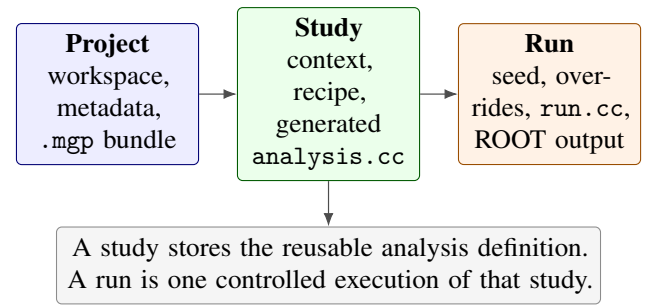

\begin{minipage}{\linewidth}
\centering
\begin{tikzpicture}[node distance=5mm]
\node[mageblue, text width=0.24\linewidth] (project) {\textbf{Project}\\workspace, metadata, \mgp{} bundle};
\node[magegreen, text width=0.24\linewidth, right=5mm of project] (study) {\textbf{Study}\\context, recipe, generated \code{analysis.cc}};
\node[mageorange, text width=0.24\linewidth, right=5mm of study] (run) {\textbf{Run}\\seed, overrides, \code{run.cc}, ROOT output};
\draw[magearrow] (project) -- (study);
\draw[magearrow] (study) -- (run);
\node[magegray, text width=0.78\linewidth, below=6mm of study] (note) {A study stores the reusable analysis definition. A run is one controlled execution of that study.};
\draw[magearrow] (study) -- (note);
\end{tikzpicture}
\captionof{figure}{Project--study--run model used by MAGE-HEP. The study maintains the reusable context, while each run stores a concrete execution of it.}
\label{fig:project-study-run}
\end{minipage}
\end{center}

\subsection{Project Bundle}

In the beta version, MAGE-HEP produces a lightweight \texttt{.mgp} bundle that compresses and stores project metadata, analysis flows, and other details in a fixed project layout. A representative project structure is as follows:

\begin{lstlisting}[
style=papercode,
caption={Representative layout of a lightweight \texttt{.mgp} project bundle.},
label={lst:mgp-bundle}
]
~/MAGE/Projects/<ProjectName>/
    project_manifest.json
    <ProjectName>.mgp

    studies/
        <study_id>/
            study_manifest.json
            context.mcx
            analysis.cc
            analysis.h
            pythia.cmnd
            observables.json
            root_layout.json

            runs/
                <run_id>/
                    run_manifest.json
                    run.cc
                    status.json
                    run.log
                    spectra.root
\end{lstlisting}

The study directory contains the reusable analysis definition and generated study-level source code \texttt{analysis.cc}, which is generated using the MAGE-HEP Node API, described in Sec.~\ref{sec:NodeAPI}. The run directory contains the files specific to one execution. This allows the user to inspect the generated code while keeping the run-level output separate.

\subsection{Execution Flow}

Algorithm~\ref{alg:mage-execution} summarizes the MAGE-HEP project creation and run-execution workflow, where \texttt{mage-daemon} executes runs in the background.

\begin{algorithm}[h]
\caption{MAGE-HEP run-execution workflow}
\label{alg:mage-execution}
\begin{algorithmic}[1]
\State Create or open a MAGE-HEP project.
\State Define or import a study context.
\State Generate the study-level \texttt{analysis.cc}.
\State Select run-level parameters exposed by the study.
\For{each requested run}
    \State Create a run directory.
    \State Write \texttt{run\_manifest.json} with seed, parameters, and output paths.
    \State Generate or configure the run-level \texttt{run.cc}.
    \State Submit the run to \texttt{mage-daemon}.
    \While{run is active}
        \State Update \texttt{status.json}.
        \State Stream log messages to the GUI.
        \State Update the progress bar.
    \EndWhile
    \State Register ROOT output and metadata.
    \State Update ROOT plots and particle tables in the GUI.
\EndFor
\State Export the lightweight project state as a \texttt{.mgp} bundle.
\end{algorithmic}
\end{algorithm}

\section{MAGE-HEP Node API}
\label{sec:NodeAPI}

The MAGE-HEP Node API is the user-side library that allows users to describe analyses, which are then converted into executable code. In a typical Monte Carlo analysis, the user writes the generator setup, event loop, particle loop, cuts, histogram definitions, ROOT output names, and run commands in a single script. Although flexible, such scripts become increasingly difficult to maintain and reproduce after multiple modifications. The purpose of the MAGE-HEP Node API is to separate this physics intent from the low-level implementation.

In MAGE-HEP, the user does not have to write the full code for each analysis; instead, they define the \textit{study context}. The user-defined study context contains a library of generator configurations, analysis modules, observables used in an analysis, and output rules, each specifying how each observable should be stored in ROOT. MAGE-HEP then builds an internal context from this information. This context is the central object of the Node API. It records what the user wants to calculate, which variables are used, where they are produced, how they are connected, and how the final output should be written.

\subsection{Internal Context Graph}

The internal context graph, which shows the node's reusability, is shown in Fig.~\ref{fig:flow}. The PYTHIA8 block defines the event source, the run node serves as a bridge to the connection for a specific analysis that uses context-based variables, the event node defines event-level quantities that can be calculated outside the particle loop or with their own particle loop, such as multiplicity or centrality, the particle node defines particle-level observables such as transverse momentum ($\pT$), pseudorapidity ($\eta$), rapidity ($y$), and azimuthal angle ($\phi$), the cut node and property blocks define selections and fill operations, the data socket stores references to the node points that can be stored, and the ROOT builder defines the code based on the user-defined output rules. Fig.~\ref{fig:flow} shows three different analyses derived from the same context. In the current version of MAGE-HEP, the user can define a context based on the C++ library of the MAGE-HEP Node API.

\subsection{User-Defined Context Example}

Listing~\ref{lst:mage-node-api-code} shows an example of user-side code written in \texttt{pp\_context.cpp}, which creates the study context for the proton-proton analysis. The code specifies the required external tools, defines a PYTHIA8 source, sets the analysis recipe, and instructs MAGE-HEP to generate the corresponding C++/ROOT analysis file and export the study context.

\begin{lstlisting}[
style=papercode,
language=C++,
caption={Study-level use of the MAGE-HEP Node API corresponding to the workflow in Fig.~\ref{fig:flow}.},
label={lst:mage-node-api-code}
]
#include "MageAPI.h"

using namespace Mage;

int main() {
    Study pp("flow_context_pp");

    pp.Toolchain()
        .Require("ROOT")
        .Require("PYTHIA8")
        .AutoDetect();

    pp.Pythia8("pp_source", {
        .system = "pp",
        .sqrtS = 13.6_TeV,
        .events = 100000000,
        .cmndFile = "pp_base.cmnd",
        .settings = {
            {"Tune:pp", 14},
            {"SoftQCD:inelastic", true}
        }
    });

    pp.Analyze(Recipes::UserHistogram{
        .name = "azimuth_analysis",
        .stage = Stage::Particle,
        .value = Particle::Phi,
        .selection = {
            Select::Final(),
            Select::Charged(),
            Cut(Expression("abs(.eta()) < etaMax"))
        },
        .parameters = {
            {"etaMax", 0.8}
        },
        .binning = {0.0, 6.28318, 0.0628318},
        .store = "phiData",
        .output = {"pp_phi.root"}
    });

    pp.Generate("generated/flow_context_pp");
    pp.Export("flow_context_pp.mcx");

    return 0;
}
\end{lstlisting}

\begin{figure*}[p]
    \centering
    \includegraphics[width=0.85\linewidth]{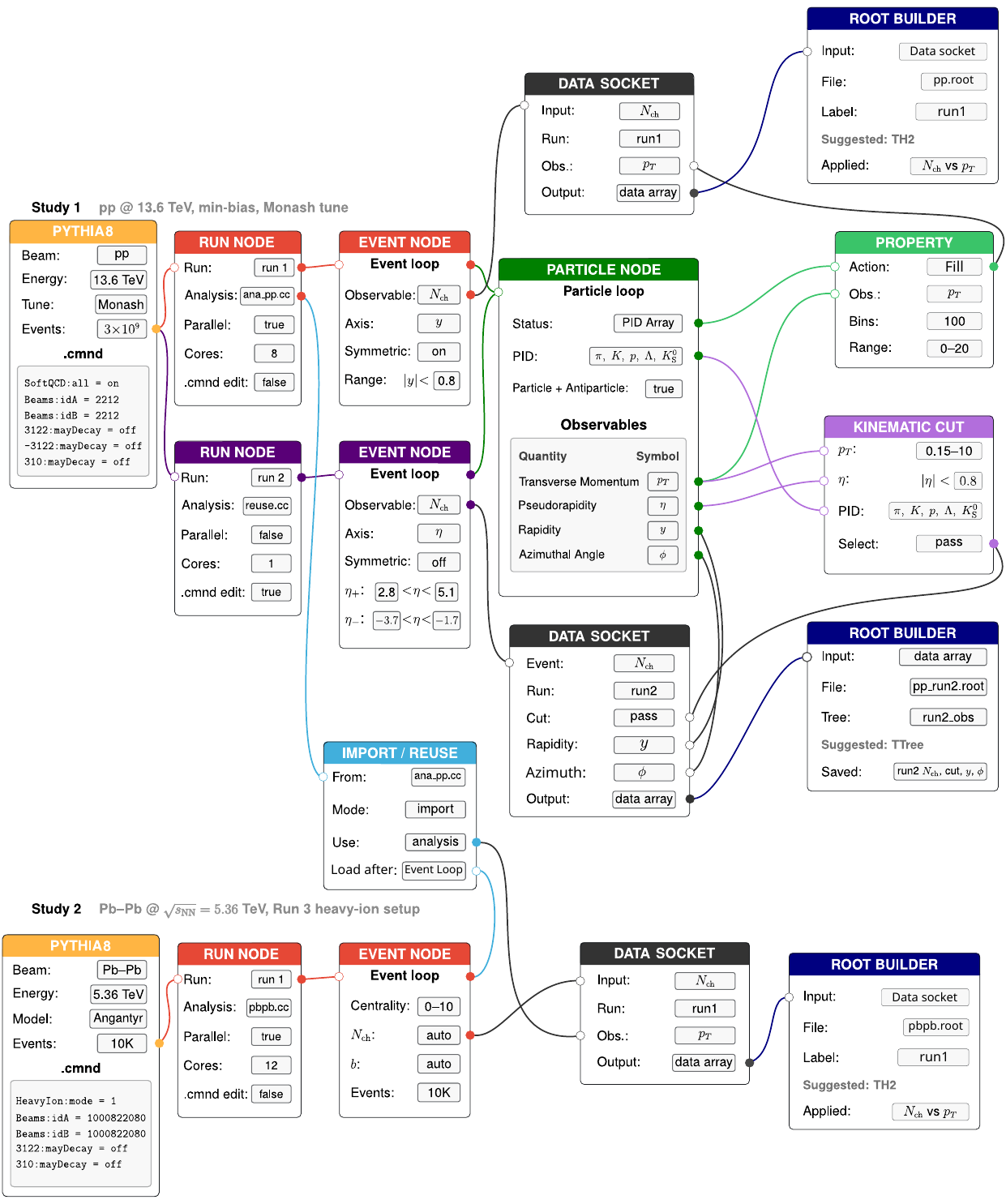}
    \caption{Conceptual context graph used by MAGE-HEP. The figure shows how event-source nodes, run nodes, event-level quantities, particle-level observables, cuts, data sockets, and ROOT builders are connected and reused across analysis branches.}
    \label{fig:flow}
\end{figure*}

Listing~\ref{lst:mage-node-api-code} represents the code of the upper \texttt{run1} block of Fig.~\ref{fig:flow}. The line \texttt{Study pp("flow\_context\_pp")} creates the analysis context. Here, the \texttt{Toolchain()} block records the required tools, and \texttt{.AutoDetect()} saves the versions of the tools used in the context file. This is important for reproducibility because the generated analysis is not only the source file; it also carries information about the tools required to rebuild it.

The \texttt{.Pythia8(...)} block corresponds to the PYTHIA8 node in the figure. It stores the collision system, center-of-mass energy, number of events, and additional PYTHIA8 settings. Common Monte Carlo generator settings can be added to user-end analysis scripts as reusable sections using \texttt{.Set(...)}, or the user can load a \texttt{.cmnd} file in the case of PYTHIA using \texttt{.LoadCmndFile("config.cmnd")}. This allows users to define generator configurations for each analysis and easily connect them to other nodes built for different analyses.

The \texttt{.Analyze(...)} block corresponds to the event, particle cuts, property, and data-socket part in Fig.~\ref{fig:flow}. In this example, the \texttt{Recipes} block tells MAGE-HEP that the observable is the azimuthal angle ($\phi$), which is evaluated at the particle stage, and that only final charged particles satisfying $|\eta|<0.8$ are selected. In the current implementation, selection conditions are represented using string-based expressions in the block \texttt{"abs(.eta()) < etaMax"}, where \texttt{.eta()} is directly parsed as \texttt{particle.eta()}, as used in the particle loop in PYTHIA8. The output of this operation is accumulated in the store under the user-defined label \texttt{phiData}. The store is the internal data socket of the analysis: it is the object that later connects the particle-level calculation to ROOT output, which is saved as \texttt{"pp\_phi.root"}.

The \texttt{.Generate(...)} call invokes the MAGE-HEP transpiler, a source-to-source translation layer~\cite{compilerbook}, to convert the context into an ordinary C++ analysis file. The generated directory contains the source code needed to compile and run the analysis, along with metadata files such as context summaries, parameter values, semantic flow information, and a ROOT layout description. It allows the user to inspect the generated code rather than treating MAGE-HEP as a black box, while also allowing the generated files to be copied, compiled, and run independently when required.

\texttt{.Export(...)} saves the analysis context as a MAGE-HEP context bundle, which is saved with the \texttt{.mcx} extension. The \texttt{.mcx} file is the reusable part of the analysis-building workflow. It stores the analysis definitions, selected modules, parameters, stores, output rules, and other important information. This makes it possible to reopen the same context later, import it into another system, or share the analysis logic without copying a long, monolithic codebase. The bundle features for the current version are limited, but the core motivation is to allow a user to add more context to a single package, treating it as the central hub of all the analysis tools and using combinations of the defined tools by editing the values of a few parameters to create a new analysis.

\subsection{Some Predefined Recipe}

The Node API also supports predefined analysis recipes for common generator-level studies. For example, the identified-particle transverse-momentum spectra module is as follows:
\begin{lstlisting}[
style=papercode,
language=C++,
caption={Use of a predefined recipe for identified-particle spectra.},
label={lst:mage-pid-pt}
]
Study pp("pid_spectra_pp");

pp.Pythia8("pp_source", {
    .system = "pp",
    .sqrtS = 13.6_TeV,
    .events = 100_kEvents,
    .cmndFile = "pp_base.cmnd"
});

pp.Analyze(Recipes::PidPtVsNch{
    .name = "pid_pt_vs_nch",
    .yMaxGlobal = 0.8,
    .etaMax = 3.0,
    .nchBinWidth = 10,
    .eta = {0.0, 3.0, 0.1},
    .pt = {0.0, 10.0, 0.1},
    .ptMin = 0.05,
    .targetV0 = {310, 3122},
    .output = {"pid_pt_spectra.root"}
});

pp.Generate("generated/pid_spectra_pp");
pp.Export("pid_spectra_pp.mcx");
\end{lstlisting}
Here, the user defines the physics requirements: the rapidity for $\Nch$ specified by \texttt{.yMaxGlobal}, the pseudorapidity acceptance cuts specified by \texttt{.etaMax}, the $\pT$ range, the multiplicity-bin width, the particle species, and the output file. Internally, MAGE-HEP expands this recipe into a particle-stage observable, creates sparse stores for the spectra, applies the event and particle logic, and writes ROOT histograms according to the output rule. For this recipe, the output rule uses $\pT$ and $\Nch$ as histogram axes and splits the result by PID and $\eta$ bin. Thus, the output is a collection of two-dimensional $\pT$ vs. $\Nch$ histograms, each carrying the corresponding particle and $\eta$ information.

The corresponding save-rule structure is discussed in Sec.~\ref{sec:root-rules}.

\subsection{Context Reuse}

In addition to the definition of the rules and variables, another important feature of the Node API is context reuse, as shown in the lower part of Fig.~\ref{fig:flow}. A previously prepared analysis can be imported and reused in a new study. This can be done by loading the existing \texttt{.mcx} and reusing one of its modules, as shown in Listing~\ref{lst:mage-reuse}.
\begin{lstlisting}[
style=papercode,
language=C++,
caption={Reuse of a previously exported MAGE-HEP context.},
label={lst:mage-reuse}
]
Study pbpb("reused_heavy_ion_context");

pbpb.Pythia8("pbpb_source", {
    .system = "PbPb",
    .sqrtSNN = 5.36_TeV,
    .events = 10_kEvents,
    .cmndFile = "pbpb_base.cmnd"
});

pbpb.Reuse("flow_context_pp.mcx");

pbpb.Generate("generated/reused_heavy_ion_context");
pbpb.Export("reused_heavy_ion_context.mcx");
\end{lstlisting}

When a context is reused through \texttt{.Reuse}, MAGE-HEP carries the module definition, parameters, store names, output rules, source-level information, and version metadata together. This is useful when the same analysis logic needs to be applied to a different collision system, generator setting, additional store, or output convention. In Fig.~\ref{fig:flow}, this corresponds to the import/reuse block, which connects the previously defined analysis to a new run branch while preserving a clean record of the analysis definitions.

\section{ROOT Builder and \textit{save-rule}s}
\label{sec:root-rules}

The ROOT builder abstracts the output-definition layer while preserving user control over output organization. MAGE-HEP allows the user to write the rules for saving the output. A \textit{save-rule} tells MAGE-HEP which store should be written, which variables should be used as histogram axes, which variables should split the output into histograms, and how the ROOT objects should be named.

For example, in the predefined recipe, the internal store contains entries that can be indexed by transverse momentum, global charged multiplicity bin, particle ID, and pseudorapidity bin. The ROOT rule for this can be summarized as:

\begin{lstlisting}[
style=papercode,
caption={Representative ROOT \textit{save rule} used by the identified-particle spectra recipe. One store is written into multiple two-dimensional ROOT histograms by splitting over particle ID and pseudorapidity bin.},
label={lst:root-rule-text}
]
Store: ptData
Axes:  pt, globalNchBin
Split: pid, etaBin
Type:  TH2D
Name:  h2_pt_globalNch_pid{pid}_etaBin{etaBin}
\end{lstlisting}

The beta version of MAGE-HEP provides two predefined ROOT save-rule classes, while more flexible user-defined \textit{save-rule} definitions are planned for later versions. The first predefined rule is a one-dimensional histogram rule for simple user-defined observables, such as azimuthal-angle distributions, implemented via \texttt{Recipes::UserHistogram}. This rule maps a single observable and its binning definition to a \texttt{TH1D} object. The second is a recipe-specific two-dimensional histogram rule. In \texttt{Recipes::PidPtVsNch}, the logical axes are defined by the recipe as transverse momentum and charged-particle multiplicity, while the generated output is internally split by particle species and pseudorapidity bin to produce a collection of \texttt{TH2D} objects. In later versions, users will be able to define ROOT \textit{save-rule}s more flexibly by specifying the axes, split variables, naming patterns, and output layout directly.

Thus, the current output layer is intentionally designed to validate \texttt{TH1D} and \texttt{TH2D} layouts. A fully general save-rule interface, where the user explicitly chooses arbitrary axes, split variables, ROOT object types, and naming patterns, is planned for later versions.

\section{Graphical Interface and Background Execution}
\label{sec:gui}

The graphical interface is the primary user-facing component of MAGE-HEP. It allows the user to create projects, add studies, inspect generated files, launch runs, and monitor output. The GUI functions as both a visualization layer and a workflow manager, connecting projects, studies, runs, generated code, \texttt{mage-daemon} execution, and output viewing.

\subsection{Project Management and Study Creation}

\begin{figure}[H]
    \centering
    \includegraphics[width=\linewidth]{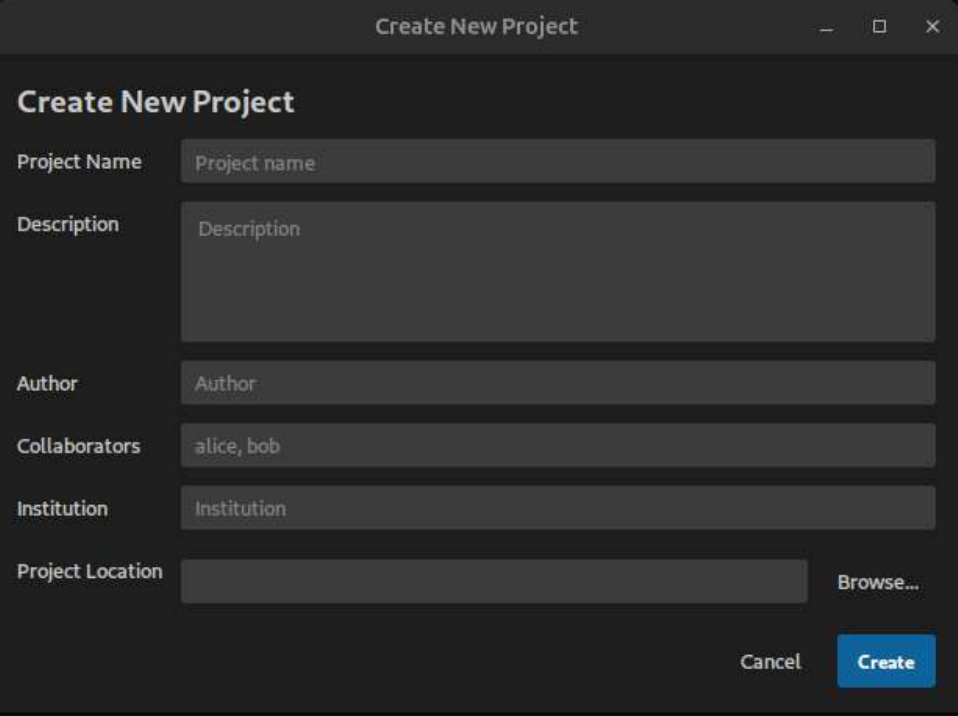}
    \caption{\texttt{Create New Project} window in the MAGE-HEP GUI. The user enters the project details and initializes a new project workspace.}
    \label{fig:createproject}
\end{figure}

A MAGE-HEP project is created from the GUI by clicking the \texttt{Create Project} button. This opens a window where the user can enter the project details, as shown in Fig.~\ref{fig:createproject}. After the project is created, it can be exported as a \texttt{.mgp} bundle.

The user can then create a study by clicking the \texttt{Add Study} button. The study-creation window allows the user to enter a basic description of the study and select the access mode for the context. If the user selects \texttt{Public}, no encryption is applied. If the user selects \texttt{Private}, the context is locked to the application-level user UUID (Universally Unique Identifier), which MAGE-HEP creates during installation and stores in an encrypted form. The \texttt{Key-gated} mode is planned for future versions, allowing users to share a context file only with selected collaborators. This can be useful for analysis definitions that are not yet ready to be made public.

\begin{figure}[H]
    \centering
    \includegraphics[width=\linewidth]{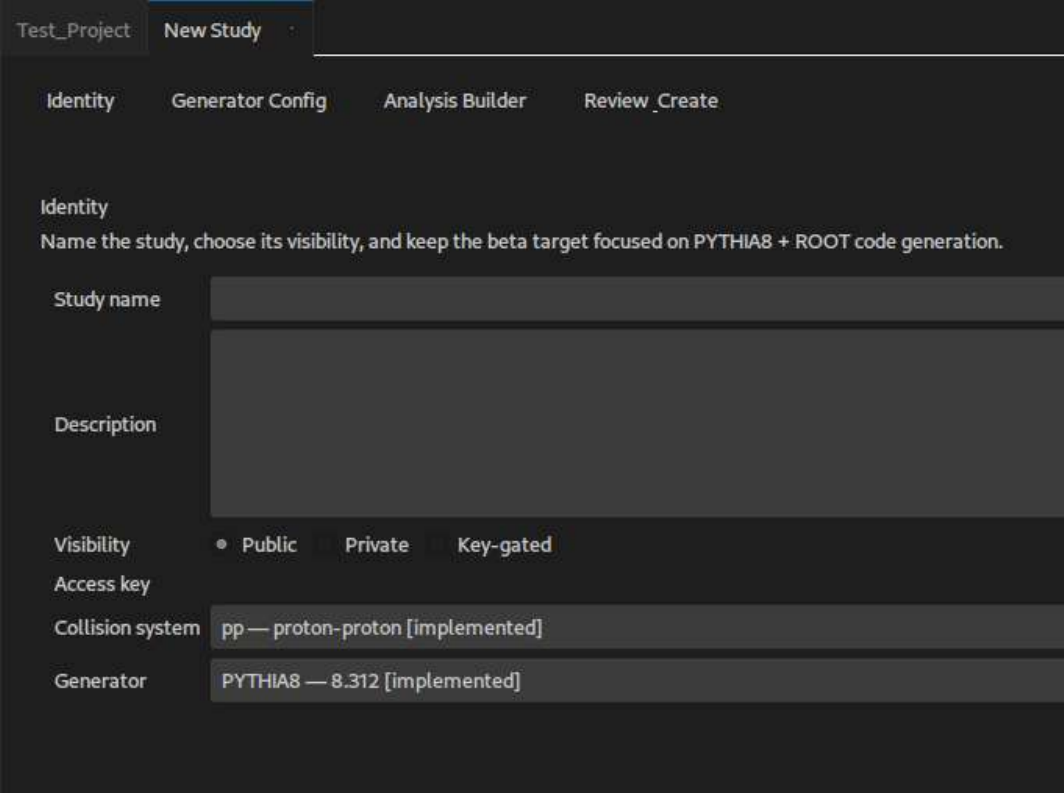}
    \caption{\texttt{New Study} window in the MAGE-HEP GUI. The user selects the
    collision system, generator configuration, and context access mode.}
    \label{fig:createstudy}
\end{figure}

As shown in Fig.~\ref{fig:createstudy}, the user can select the collision system, generator, and generator configuration. In the \texttt{Analysis Builder} panel, the user can either upload a context definition file to create a \texttt{.mcx} context bundle, or upload an existing \texttt{.mcx} file and load it into the internal MAGE-HEP context. MAGE-HEP then generates the study-level \texttt{analysis.cc} and the run-level \texttt{run.cc} files.

\subsection{Run Overview}

\begin{figure}[!htbp]
    \centering
    \includegraphics[width=\linewidth]{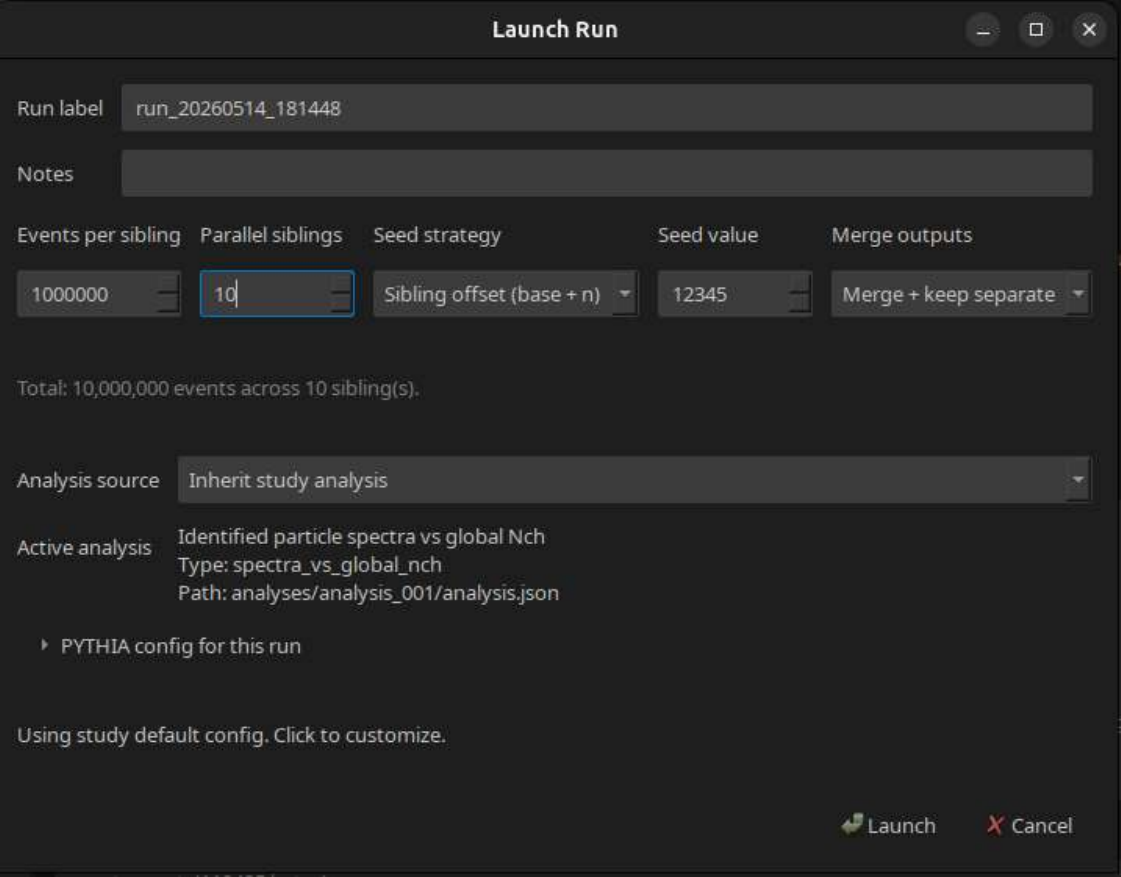}
    \caption{\texttt{Launch Run} window in MAGE-HEP GUI. The user can modify exposed run-level parameters and select the run output policy.}
    \label{fig:createrun}
\end{figure}

\begin{figure*}[!htbp]
    \centering
    \includegraphics[width=\linewidth]{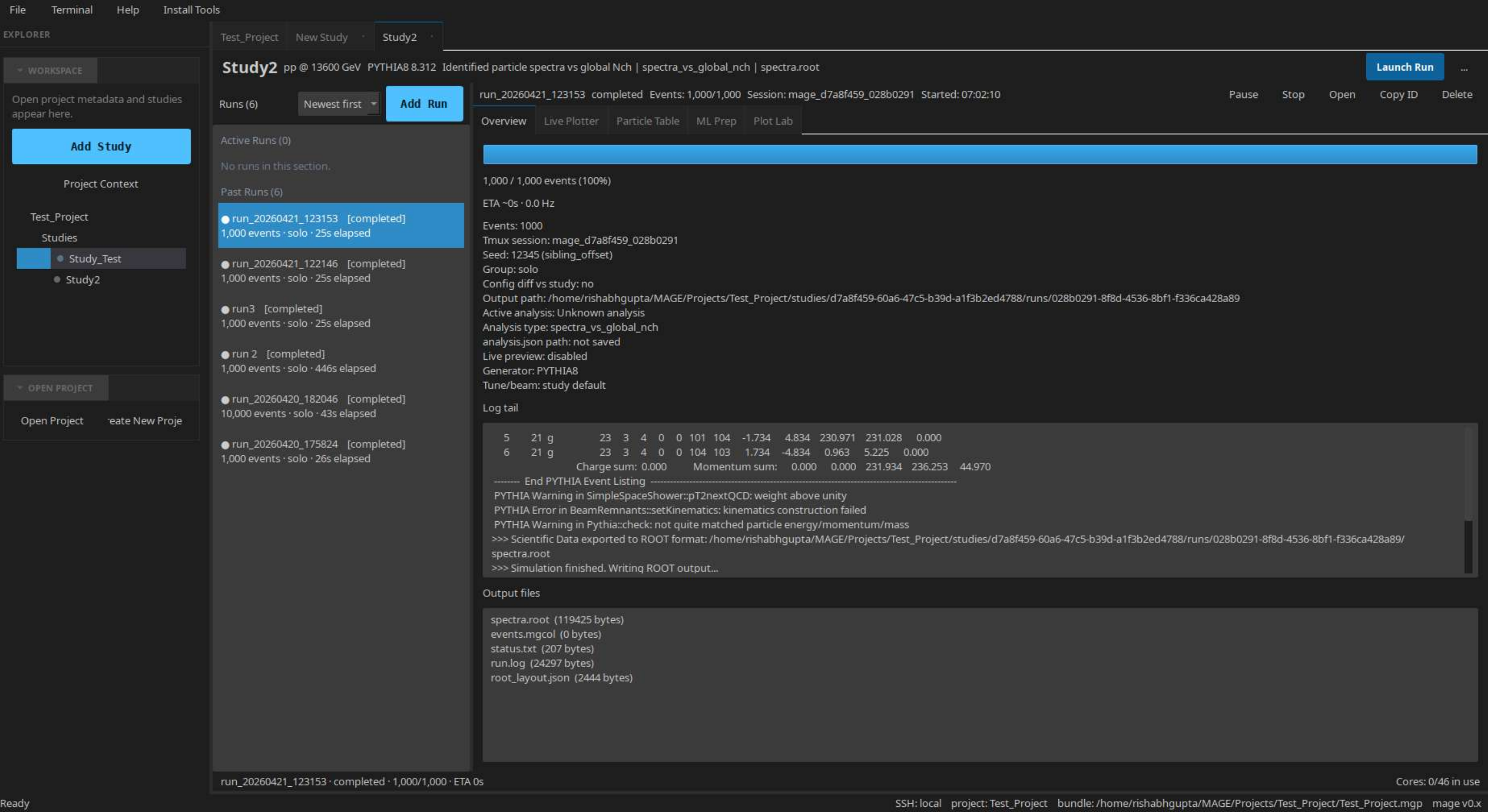}
    \caption{Run overview panel in MAGE-HEP. The interface displays the run
    registry, progress information, generated files, and status updates.}
    \label{fig:runoverview}
\end{figure*}

After a study is created, the user can add a run by clicking the \texttt{Add Run} button, as shown in Fig.~\ref{fig:runoverview}. The run window allows the user to modify the values of the study context variables and execute the run. The user can execute either a single run or multiple runs by selecting the number of \texttt{Parallel siblings}, as shown in Fig.~\ref{fig:createrun}. In this context, siblings denote parallel run instances that are executed as controlled variations of the same study context.

MAGE-HEP also allows the user to choose the output policy through the \texttt{Merge outputs} dropdown. Each run may produce a separate ROOT output file, or outputs from multiple runs may be merged into a single ROOT file, depending on the selected configuration. An overview of the run interface is shown in Fig.~\ref{fig:runoverview}. \texttt{Overview} tab contains the progress bar and the registry of all runs. The progress of each run is updated by \texttt{mage-daemon}.

\subsection{Background Execution with \texttt{mage-daemon}}

To prevent the graphical interface from slowing down during long runs, MAGE-HEP uses \texttt{mage-daemon} to execute runs in the background. The daemon receives run requests from the GUI, launches the generated run driver in a detached background terminal, tracks the process, updates the status file, writes logs, and reports progress back to the interface.

This design allows several runs to be executed in parallel using the available CPU cores. Each run has its own status and log files, allowing the project state to be recovered even if the GUI is closed and reopened.

\subsection{Live ROOT Plots and Particle Table}

\begin{figure*}[!htbp]
    \centering
    \includegraphics[width=\linewidth]{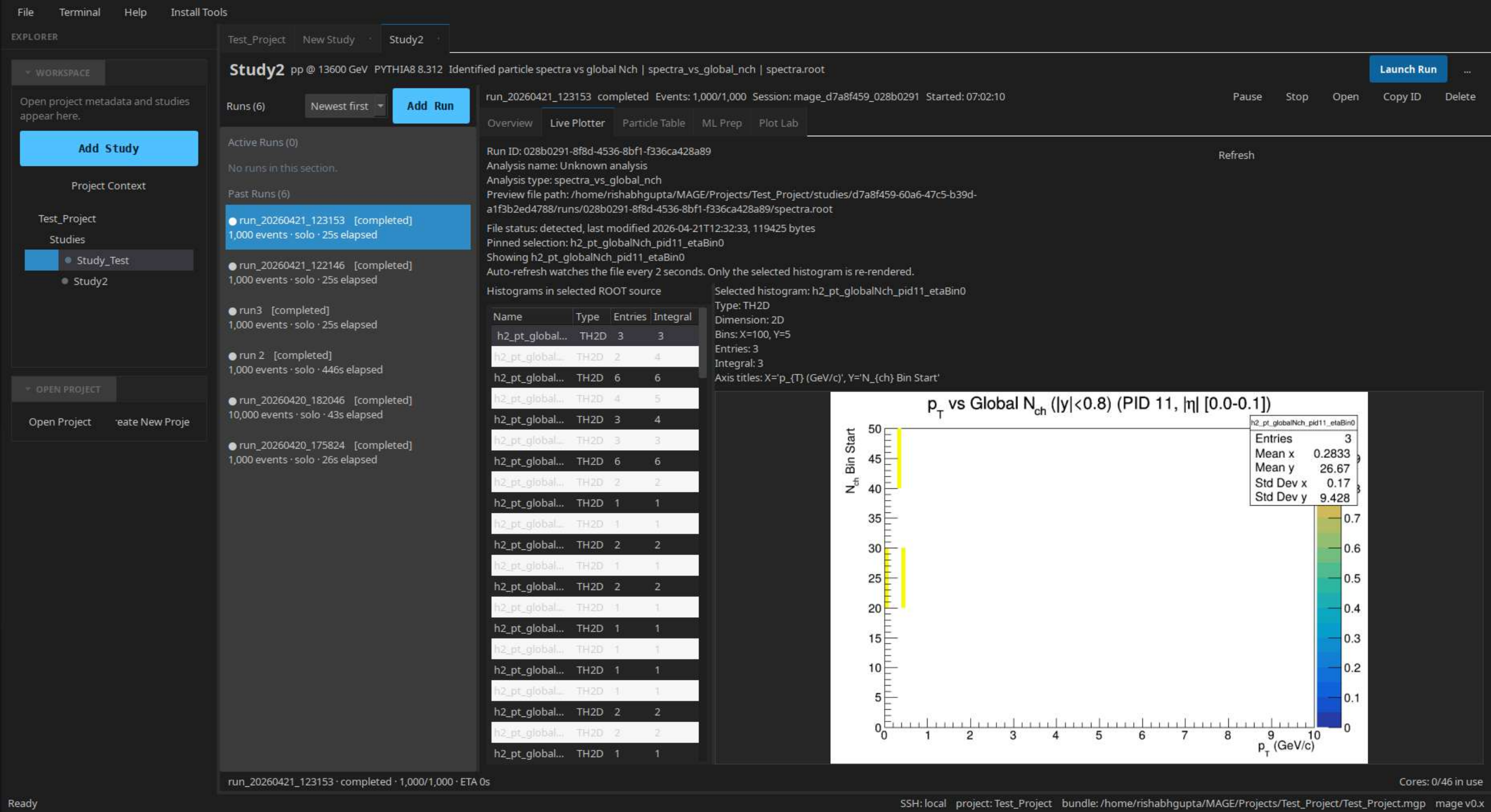}
    \caption{Live ROOT plotting panel in MAGE-HEP GUI. }
    \label{fig:liveroot}
\end{figure*}

The ROOT \textit{save-rule}s allow MAGE-HEP to track which variables are saved during a run. Therefore, the GUI can display ROOT histograms directly inside the MAGE-HEP interface. This allows users to inspect selected plots during or after a run without manually opening ROOT for every output file, as shown in Fig.~\ref{fig:liveroot}.

\begin{figure*}[!htbp]
    \centering
    \includegraphics[width=\linewidth]{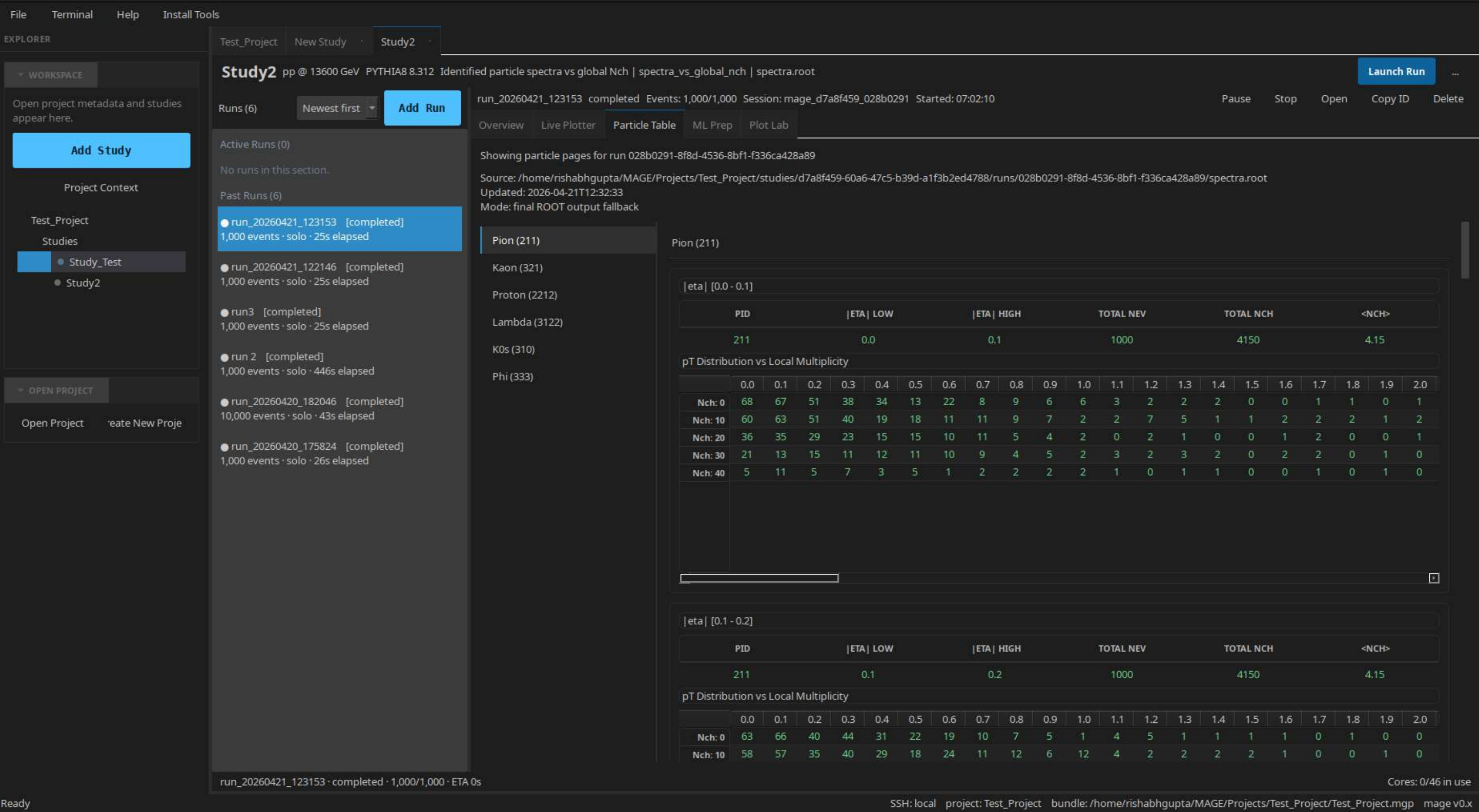}
    \caption{Particle table view in MAGE-HEP GUI.}
    \label{fig:liverptable}
\end{figure*}

The GUI can also display particle-level summary tables for supported analysis formats. In the current version, this view is intentionally restrictive and is mainly used to validate identified-particle spectra workflows, as shown in Fig.~\ref{fig:liverptable}. Both tools follow the lazy-loading principle, where output objects are loaded only when needed, helping the GUI handle large output files without unnecessary memory overhead \cite{fowler2003patterns}.

In the current version, the GUI is restricted to context creation, run management, project management, and basic visual aids. Future versions will extend this interface to support broader usage. Features such as \texttt{ML Lab} and \texttt{Plot Lab} are also planned. The \texttt{ML Lab} will allow users to build an ML dataset directly from ROOT files after a run, based on user-defined requirements, without manually loading each file and generating the dataset separately. In the current version, this feature is restricted to the dataset-generation workflow used in Ref.~\cite{Gupta:2026gzt}. The \texttt{Plot Lab} is planned to allow users to plot derived quantities directly from ROOT files using the stored context, without writing additional analysis code.

\section{Preview of MAGE-HEP Output}
\label{sec:output}

As a representative example, we show the output produced by the predefined recipe in Listing~\ref{lst:mage-pid-pt}. In this workflow, the user defines a PYTHIA8 source, a study context, and an identified particle transverse-momentum spectra recipe. The recipe computes particle spectra as a function of $\pT$ and $\Nch$. The output is written to ROOT according to the corresponding \textit{save-rule}.

\begin{figure}[H]
    \centering
    \includegraphics[width=\linewidth]{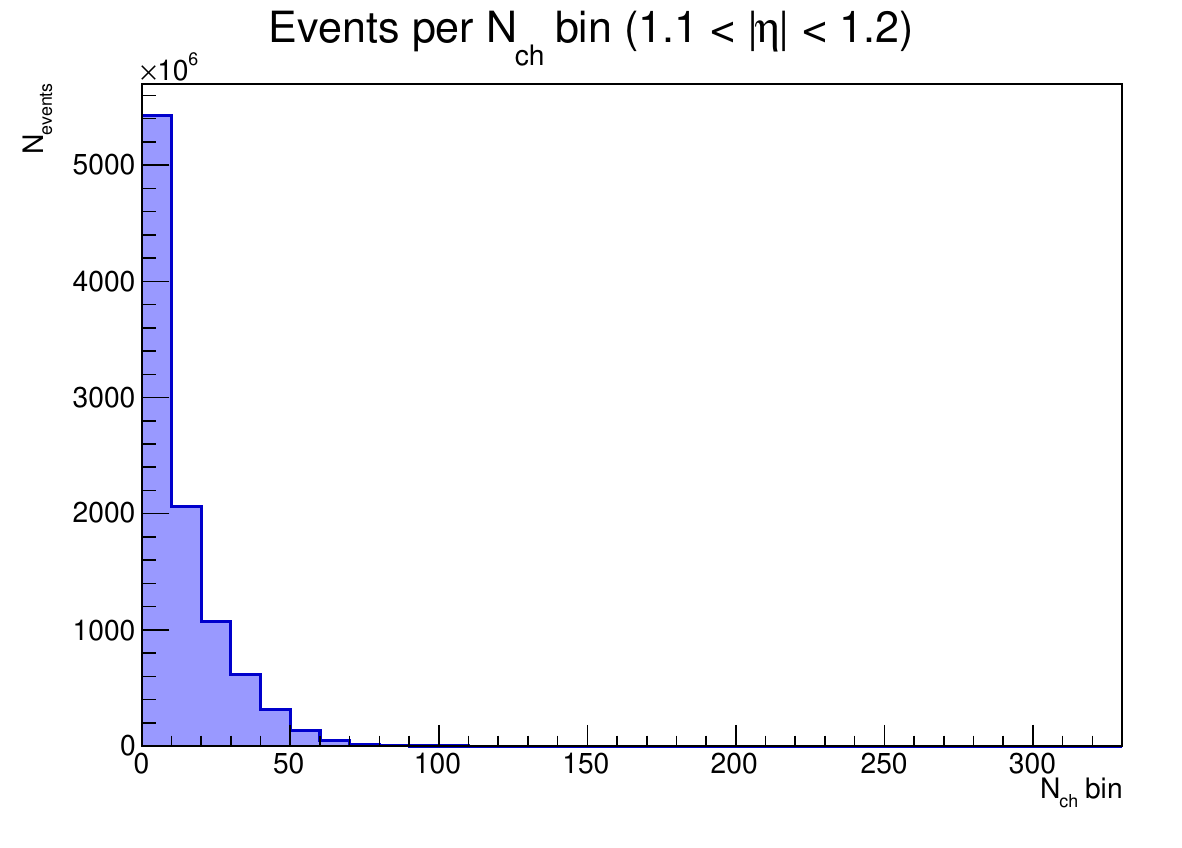}

    \vspace{0.3cm}

    \includegraphics[width=\linewidth]{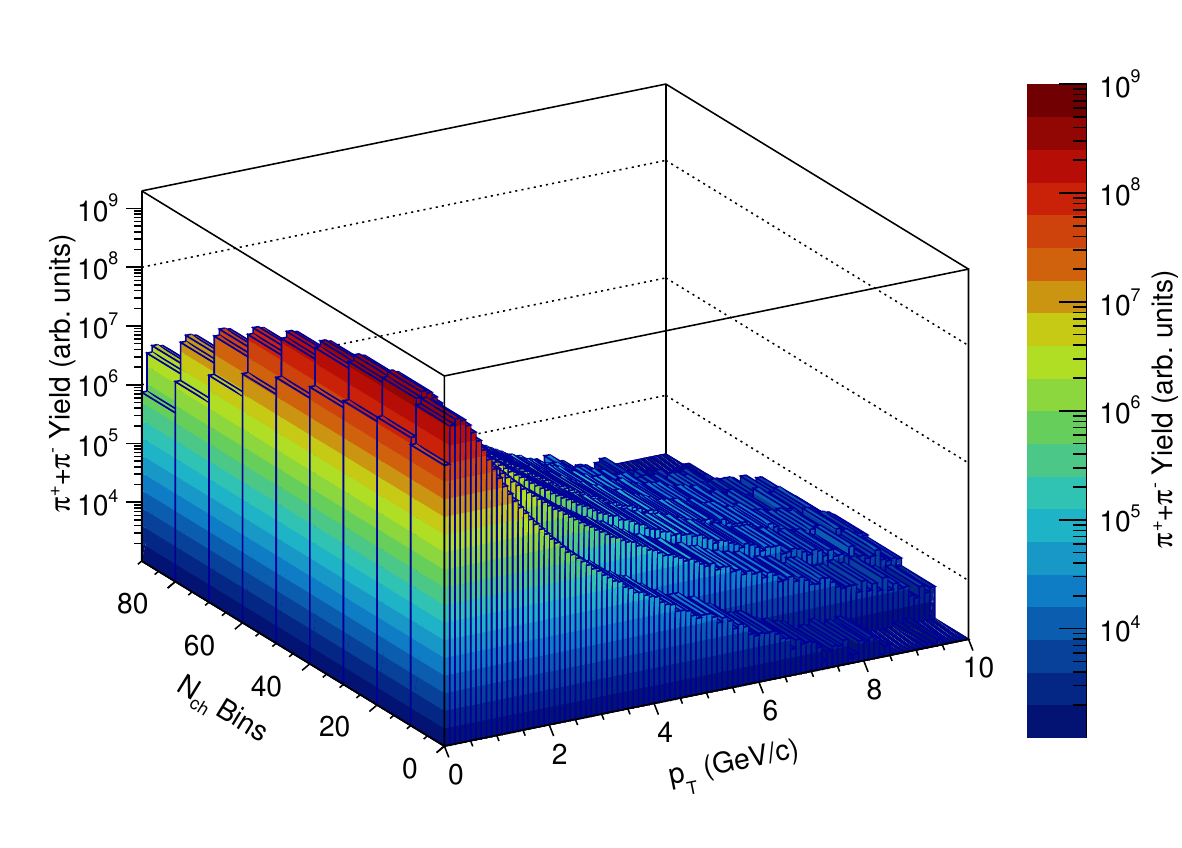}

    \caption{Example output generated from the predefined identified-particle spectra recipe \texttt{Recipes::PidPtVsNch}. The upper panel shows the event distribution in global charged-multiplicity bins for \(1.1 < |\eta| < 1.2\). The lower panel shows the transverse-momentum distribution as a function of charged multiplicity for the selected particle and pseudorapidity bin.}
    \label{fig:results}
\end{figure}

The generated context was executed for approximately \(10^{10}\) events, distributed across 40 sibling cores. The run outputs were then merged into a single ROOT file. In this example, the output layer is focused on the $\pT$ and $\Nch$ structure of the predefined recipe. The resulting distributions are shown in Fig.~\ref{fig:results}. The upper panel shows the event distribution in global charged-multiplicity bins for the selected pseudorapidity interval. The lower panel shows the generated $\pT$ versus $\Nch$ output for the selected particle and pseudorapidity bin.

This example validates the implementation of the MAGE-HEP Node API by demonstrating that the analysis definition generated via the API is successfully translated into C++/ROOT analysis files and produces the expected ROOT outputs and plots. The analysis logic is defined once at the study level, while individual runs represent controlled executions of the same context. The generated C++/ROOT code remains inspectable and can be run independently, while the GUI maintains execution metadata linked to the generated ROOT outputs.

\section{Conclusion}
\label{sec:conclusion}

MAGE-HEP is a GUI-based analysis workflow for reproducible high-energy physics Monte Carlo studies. It allows users to configure and execute Monte Carlo simulations without manually managing the full analysis workflow. MAGE-HEP introduces a project-study-run structure, where reusable analysis definitions are stored at the study level, and individual runs are executed as controlled variations of that context. This separation allows the generator configuration, analysis logic, generated code, run parameters, logs, ROOT outputs, and tool versions to be tracked together.

As part of MAGE-HEP, we introduce the MAGE-HEP Node API as the analysis-building layer. It allows users to define generator sources, observable selections, stores, and ROOT output rules in a structured form. The transpiler converts this context into inspectable C++/ROOT analysis code that can also be compiled and run independently outside the graphical interface. The graphical interface then manages project creation, run execution, progress monitoring, live output inspection, and export of the workflow state.

The current beta implementation demonstrates the main concepts of MAGE: GUI-based project creation, reusable study contexts, generated C++ analysis files, background execution through \texttt{mage-daemon}, ROOT output production, live plot inspection, particle-table summaries for supported formats, and the other workflow features described earlier. The present implementation is restricted to the validated PYTHIA8 and ROOT workflow, selected ROOT output layouts, and a limited set of analysis recipes. Fully flexible user-defined \textit{save-rule}s, additional generator backends, and broader output formats are planned for later versions.

The MAGE-HEP source code is currently being validated for public release. A public release, including installation instructions, example studies, and sample \texttt{.mgp} bundles and reproducibility files are planned after validation of a stable version. Future work will focus on stabilizing the architecture, expanding the supported ROOT output layouts, improving the project bundle format, extending the Node API, adding more generator backends, and providing validated example studies with installation and reproducibility instructions.

\section*{Acknowledgment}
K.G. gratefully acknowledges the support from the Prime Minister's Research Fellowship (PMRF), Government of India. S.P. acknowledges the financial support from the Hungarian National Research, Development and Innovation Office (NKFIH) under the contract numbers NKFIH NKKP ADVANCED\_25-153456, 2025-1.1.5-NEMZ\_KI-2025-00005, 2024-1.2.5-TET-2024-00022, and the usage of Wigner Scientific Computing Laboratory (WSCLAB). The authors (K.G., and R.S.) acknowledge funding from the DAE-DST, Government of India, under the mega-science project “Indian participation in the ALICE experiment at CERN,” bearing Project No. SR/MF/PS-02/2021-IITI(E-37123). This work has been carried out as a part of the M.Sc. (Physics) thesis by Rishabh Gupta at IIT Indore.

\end{document}